\documentclass[prd,showpacs,showkeys,eqsecnum,nofootinbib,preprint,superscriptaddress]{revtex4-1}
\usepackage{epsfig}
\usepackage{amsmath}
\usepackage{amsfonts}
\usepackage{amssymb}
\usepackage{color}
\usepackage{graphicx}
\usepackage{url,hyperref}
\def  \bcen   {\begin{center}}
\def  \ecen   {\end{center}}
\def  \beq    {\begin{equation}}
\def  \eeq    {\end{equation}}
\def  \beqa   {\begin{eqnarray}}
\def  \eeqa   {\end{eqnarray}}

\def\bea{\begin{eqnarray}}
\def\eea{\end{eqnarray}}

\def \eegah     {$e^+e^- \to \gamma h$ }

\begin{document}
\title{Associated Production of Higgs at Linear Collider\\ 
in the Inert Higgs Doublet Model}
\author{Abdesslam Arhrib}
\email{aarhrib@ictp.it}
\affiliation{D\'{e}partement de Math\'{e}matique, 
Facult\'{e} des Sciences et Techniques,
Universit\'{e} Abdelmalek Essaadi, B. 416, Tangier, Morocco}
\affiliation{Institute of Physics, Academia Sinica, Nankang, 
Taipei 11529, Taiwan}
\author{Rachid Benbrik}
\email{rbenbrik@ictp.it}
\affiliation{D\'epartement de Physique, Facult\'e Polydisciplinaire de Safi, 
Sidi Bouzid B.P. 4162, 46000 Safi, Morocco}
\affiliation{Instituto de Fisica de Cantabria (CSIC-UC), Santander, Spain}
\affiliation{LPHEA, FSSM, Cadi Ayyad University, B.P. 2390, Marrakesh, Morocco}
\author{Tzu-Chiang Yuan}
\email{tcyuan@phys.sinica.edu.tw}
\affiliation{Institute of Physics, Academia Sinica, Nankang, 
Taipei 11529, Taiwan}

\begin{abstract}
\noindent 
We study the correlation between the Standard Model Higgs decay $h\to \gamma \gamma$ and $h\to Z \gamma$
in the Inert Higgs Doublet Model. It is found that 
these two one-loop-induced decays are positively correlated, with the latter channel having slightly
smaller branching ratio than the former one.
At the Linear Collider, we study the interplay of the off-shell extension of these two amplitudes that
contributed significantly to the associated production of the Higgs boson with a photon in the process
$e^+ e^- \to \gamma h$ and with an electron in the process
$e^-\gamma \to e^- h$ in the $s$ and $t$ channels respectively via both $\gamma$ and $Z$ exchange
for each process.

\end{abstract}
\maketitle
%%%%%%%%%%%%%%%%%%%%%%%%%%%%%%%%%%%%%%%%%%%%%%%%%%
\section{Introduction \label{section:1}}
%%#######################################################%%
Recently the ATLAS and the CMS Collaborations using the combined $7\oplus 8$
TeV data found a bosonic resonance with a mass around 125-126 GeV in 
two photons, 2 $Z$ and 2 $W$ channels  \cite{ATLAS, CMS}. 
This discovery is also confirmed by the final result from Tevatron 
 at CDF and D0 experiments through the associated production process 
 $p\bar{p}\to Wh\to (l\nu) (b\bar{b})$ \cite{tevatron}.
The new particle is necessarily a boson, since it decays into two photons, 2 $Z$ and 
2 $W$ bosons, and it could possibly be the missing particle of the 
Standard Model (SM), the Brout-Englert-Higgs boson $h$. 
For this Higgs-like particle, ATLAS obtained its mass of 
$125.5\pm 0.2(\rm{stat.})^{+0.5}_{-0.6}(\rm{syst.})$ GeV~\cite{atlas9}, while CMS got 
$125.7 \pm 0.3(\rm{stat.}) \pm 0.3(\rm{syst.})$ GeV~\cite{cms10}.
At the Moriond and EPS conferences, ATLAS and CMS updated their results 
on $h\to \gamma \gamma$, $ZZ$, $WW$, $\tau\tau$ and $b\bar{b}$ channels 
with an integrated luminosity of up to 5 fb$^{-1}$ at 7 TeV and up to 
21 fb$^{-1}$ at 8 TeV. For ATLAS the combined signal strength is found to 
be $\mu = 1.23\pm 0.18$ at the new combined mass measurement 
\cite{atlas11,Aad:2013wqa}. For the CMS update, the combined 
strength is found to be $\mu = 0.8 \pm 0.14$~\cite{cms10,Chatrchyan:2013lba}. 
All these latest experimental developments receive their greatest excitement by 
the announcement of this year 2013 Nobel Prize of Physics being awarded to F. Englert and P. W. Higgs
due to their seminal works \cite{BEH} 5 decades ago.

Since the new particle decays to pairs of gauge 
bosons and fermions, a non-integer spin is already ruled out. 
According to the Landau-Yang theorem~\cite{Landau:1948kw},
given the fact that the new boson decays into pair of photons,
it excludes the spin 1 possibility and then the remaining possibility is 
either  0 or 2. 
Recently, spin and parity of the Higgs-like particle were studied from
the angular distributions of the diphoton, $ZZ^{*}$ and $WW^{*}$ decay channels
\cite{atlaspin,cmspin,Bolognesi:2012mm} at ATLAS and CMS by looking at the
kinematical information of the final states: photons and leptons. 
Both collaborations disfavor the pure pseudoscalar or spin-2 hypothesis.
In the case of the disfavored CP-odd Higgs with $J^{PC}=0^{-+}$ its 
branching ratio into a pair of $W$s or $Z$s  is expected to be two orders of magnitude 
smaller than the observed one. From these analysis, the spin one hypothesis 
is also disfavored with an even higher confidence level.

In order to further validate the Higgs mechanism of mass generation in the SM, 
one still need to establish the following measurements with high precision: 
(1) the spin of the Higgs boson, (2) its $CP$ quantum number, (3) its couplings to 
fermions and to gauge bosons, and (4) the triple and quartic self-couplings of the
Higgs boson. 

After more than two decades of studies, technical design report for the International Linear Collider (ILC) 
has now been completed (see the Technical Design Report~\cite{tdesign:0,tdesign:V2} for details).  
Indeed, detailed simulations for various physical cases with
realistic detector properties show that the ILC can 
achieve impressive precision measurements for Higgs and top quark physics~\cite{Weiglein:2004hn,tdesign:1}. 
The ILC~\cite{tdesign:2} program will be running 
for center-of-mass energies between 200 and 500 GeV, 
with rapid changes in energy to allow for threshold scans such as 
$Zh$ at 250 GeV, $t\bar{t}$ at 350 GeV, as well as $Zhh$ and $t\bar{t}h$ at 500 GeV. 
Ultimately, increasing the ILC center-of-mass energy to 1 TeV is also envisioned. 

Clearly, first run of the LHC at $7\oplus 8$ TeV has initiated
the first step of a precise measurement program for Higgs physics 
which will get improved at the LHC 13-14 TeV run with more data 
accumulated.
It is well known that the precise measurement programs at the ILC
and LHC are complementary to each others in 
many aspects \cite{Weiglein:2004hn,Peskin:2012we}. 
Options of $\gamma\gamma$ and $e^-\gamma$ collisions 
at the ILC provide an unique opportunity for precise measurements
for Higgs properties. Thus ILC can yield substantial improvements over LHC 
measurements. Moreover, ILC will have great advantage in term of quality 
on signatures of new physics which may be overwhelmed by huge QCD backgrounds at the 
hadronic environment of LHC. 

The extraction of the Higgs-like couplings to gauge bosons
and fermions achieved up to now from the $7\oplus 8$ TeV data shows 
that this new boson behaves more and more like a SM Higgs boson 
\cite{atlas11,Aad:2013wqa,cms10,Chatrchyan:2013lba}.
More data is needed in order to fully pin down the exact 
nature of the newly discovered particle. 
The fact that the Higgs-like particle couplings
to gauge bosons and fermions are consistent with SM prediction can put severe
constraints on all models extending the SM that try to accommodate such 
Higgs-like particle.

As we now know,  the loop induced process $h\to \gamma\gamma$ turns out to be 
a discovery mode for the 125-126 GeV Higgs using the existing LHC data. 
The other related loop induced decay $h\to Z\gamma$ has not been seen yet but is expected to be measured
at the future LHC 13-14 TeV run when more data is accumulated.
Any additional charged particles beyond those in the SM will contribute to the loop amplitudes
for these two processes. Thus it is important to measure these two modes as accurate as possible.
An alternative way to extract the 
$h \gamma\gamma$ and $h Z\gamma$ couplings is to study the associated production of
$e^+e^-\to \gamma^*,Z^*\to \gamma h$ at the ILC.  Fusion production of $\gamma \gamma \to h$ 
and associated production with an electron via $e^- \gamma \to e^- h$ 
are also interesting to study if these options of $\gamma \gamma$ and $e^- \gamma$ 
collisions are available at the ILC.
 
In this paper, we concentrate on the Inert Higgs Doublet Model (IHDM) 
which is basically a two Higgs Doublet Model (THDM) with an exact $\mathbb{Z}_2$ symmetry imposed.
Under the $\mathbb{Z}_2$ symmetry, all the SM particles are even and only the second 
Higgs doublet is odd. The model was first proposed by Deshpande and Ma \cite{Deshpande:1977rw} 
to study the pattern of electroweak symmetry breaking.
Much later, it was extended further as a model of scalar dark matter together with
a radiative seesaw mechanism of neutrino mass \cite{Ma:2006km}. 

We organize this paper as follows. 
In section II, we briefly review IHDM to set up our notations and 
mention some theoretical and experimental constraints for the model.
We discuss the correlation of the signal strengths
for the two loop-induced processes $h \to \gamma \gamma$ and $h \to Z \gamma$ 
in IHDM in section III. 
We study the two processes $e^+ e^- \to \gamma h$ and $e^- \gamma \to h e^-$ in IHDM
at the ILC in section IV.
We conclude in section V.

%%#######################################################%%
\section{The Inert Higgs Doublet Model \label{section:2}}
%%#######################################################%%
Besides the SM Higgs doublet $H_1$, the IHDM \cite{Deshpande:1977rw}  
employs an additional Higgs doublet $H_2$, which can be parameterized as follows
\beq
H_1 =
   \left( \begin{array}{c}  G^+ \\ v/\sqrt{2} + 
(h + i G^0)/\sqrt{2}  \end{array}
     \right)  ~~, ~~~
H_2 =    \left( \begin{array}{c}  H^+ \\ 
(S  + i A)/\sqrt{2}  \end{array}  \right)
\eeq
where $G^\pm$ and $G^0$ are the charged and neutral goldstone bosons.
IHDM imposes a discrete $\mathbb{Z}_2$ symmetry under which all the SM fields and $H_1$ are even 
while $H_2$ is odd. The scalar potential allowed by the $\mathbb{Z}_2$ symmetry is given by
\bea
V  = \mu_1^2 |H_1|^2 + \mu_2^2 |H_2|^2 & +& \lambda_1 |H_1|^4
+ \lambda_2 |H_2|^4 +  \lambda_3 |H_1|^2 |H_2|^2 + \lambda_4
|H_1^\dagger H_2|^2 \nonumber \\
& + &\frac{\lambda_5}{2} \left\{ (H_1^\dagger H_2)^2 + {\rm h.c.} \right\} \; .
\label{potential}
\eea
The electroweak gauge symmetry is broken when $H_1$ develops 
its vacuum expectation value (VEV)
$\langle H_1 \rangle^{\rm T} = \left( 0, \; v/\sqrt{2} \right)$, 
while $\langle H_2 \rangle = 0$ to maintain the $\mathbb{Z}_2$ symmetry 
so as to allow for a dark matter (DM) candidate in this inert doublet.
This pattern of symmetry breaking results in two CP even neutral scalars ($h$, $S$), 
one CP odd neutral scalar ($A$), and a pair of charged scalars ($H^\pm$). 
Note that $h$ is the SM higgs and is $\mathbb{Z}_2$ even, while $S$, $A$ and 
$H^\pm$ are $\mathbb{Z}_2$ odd.
Only SM Higgs $h$ couples to SM fermions, 
while $S$, $A$ and $H^\pm$ are inert and do not couple to fermions.
The lighter one of the two scalars $S$ or $A$ can be a cold dark matter candidate in IHDM.
In what follows, we will denote $\chi$ as the DM candidate in this model whether it is $S$ or $A$.
Many phenomenological aspects of dark matter physics in IHDM had been studied over the years
\cite{Barbieri:2006dq,Gustafsson:2007pc,Goudelis:2013uca}. 
For an updated global analysis of IHDM, we redirect 
our readers to Ref.~\cite{Arhrib:2013ela} where extensive references of previous works can be found as well.

The masses of the 4 physical scalars can be written in terms of the parameters 
$\mu_2^2$ and $\lambda_i \, (i=1,3,4,5)$ as
\begin{eqnarray}
m_h^2 & =& -2 \mu_1^2=2 \lambda_1 v^2 \; , \nonumber \\
m_S^2 & =& \mu_2^2+\frac{1}{2} (\lambda_3 +\lambda_4 + \lambda_5) v^2
= \mu_2^2+ \lambda_L v^2 \; ,\nonumber \\
m_A^2 & = & \mu_2^2+\frac{1}{2} (\lambda_3 +\lambda_4 - \lambda_5) v^2 \; , \nonumber \\
m_{H^\pm}^2 & = &\mu_2^2+ \frac{1}{2} \lambda_3  v^2 \; , \nonumber 
\end{eqnarray}
where we have defined $\lambda_L \equiv \frac{1}{2}(\lambda_3 +\lambda_4 + \lambda_5)$
for later convenience.
One can also invert the above relations to write the quartic 
coupling $\lambda_i (i=1,3,4,5)$ in favor of the 4 physical scalar 
masses and the parameter $\mu_2^2$, 
\beq
\left\{ \lambda_1 , \lambda_3 , \lambda_4 , \lambda_5 \right\} = 
\frac{1}{v^2} \left\{ \frac{m_h^2}{2} , 2 \left( m^2_{H^\pm} - \mu^2_2 \right)  , 
\left( m_S^2 + m_A^2 - 2 m^2_{H^\pm} \right) , \left( m_S^2 - m_A^2\right)  \right\} \quad .
\label{lambds}
\eeq
In our numerical study presented in the next two sections, 
we will choose the following set of parameters
\beq
{\cal P} = \{m_h, m_S, m_A, m_{H^\pm}, \lambda_2 , \mu_2^2 \}
\label{eq:input_par_2}
\eeq
to fully describe the scalar sector of IHDM. 
%%%%%%%%%%%%%%%%%%%%%%%%%%%%%%%%%%%%%
\subsection{Theoretical and experimental constraints}
The parameter space of IHDM discussed above is subjected to both
theoretical and experimental constraints as we will describe briefly here.
\begin{itemize}
\item {Inert Vacuum:}
In order to realize the inert vacuum described earlier, one must have~\cite{Ginzburg:2010wa}:
\begin{eqnarray}
m_h^2, m_H^2, m_A^2, m_{H^\pm}^2 >0 \qquad {\rm and} \qquad 
\mu_1^2/\sqrt{\lambda_1}<
\mu_2^2/\sqrt{\lambda_2}
\end{eqnarray}
\item {Perturbativity and Unitarity:} 
Perturbativity requires all quartic couplings of the scalar potential in
Eq.~(\ref{potential}) obey $|\lambda_i| \le 8 \pi $. 
Tree-level unitarity can also be imposed by considering
a variety of scattering processes: scalar-scalar scattering,
gauge boson-gauge boson scattering and scalar-gauge boson scattering.
We impose these unitarity constraints as derived in~\cite{unitarity}.
\item {Vacuum Stability:}
To order to maintain the scalar potential $V$ bounded from below,
the following constraints on the IHDM parameters must be meet~\cite{pedro}
\begin{eqnarray}
\lambda_{1,2} > 0 \quad , \quad \lambda_3 + \lambda_4 -|\lambda_5| +
2\sqrt{\lambda_1 \lambda_2} >0 \quad\rm{and} \quad\lambda_3+2\sqrt{\lambda_1
  \lambda_2} > 0
\end{eqnarray}
\item {Experimental Constraints:}
For the experimental constraints from electroweak precision tests and collider Higgs searches,
we will follow the strategy used in~\cite{inert,Swiezewska:2012eh}.  
These constraints can be summarized as follows: 
(1) $m_{H^\pm}>80$ GeV (adapted from chargino search at LEP-II), 
(2) max$(m_A,m_{H^\pm})>100$ GeV (adapted from neutralinos
search at LEP-II), as well as (3) $m_A+m_S>m_Z$ from the $Z$ width.
\end{itemize}

%%%%%%%%%%%%%%%%%%%%%%%%%%%%%%%%%%%%%%%%%%
\section{Correlation of the signal strengths for $h \to \gamma \gamma$ and $h \to Z \gamma$}

\begin{figure}[t]
\begin{center}
\includegraphics[scale=1.03]{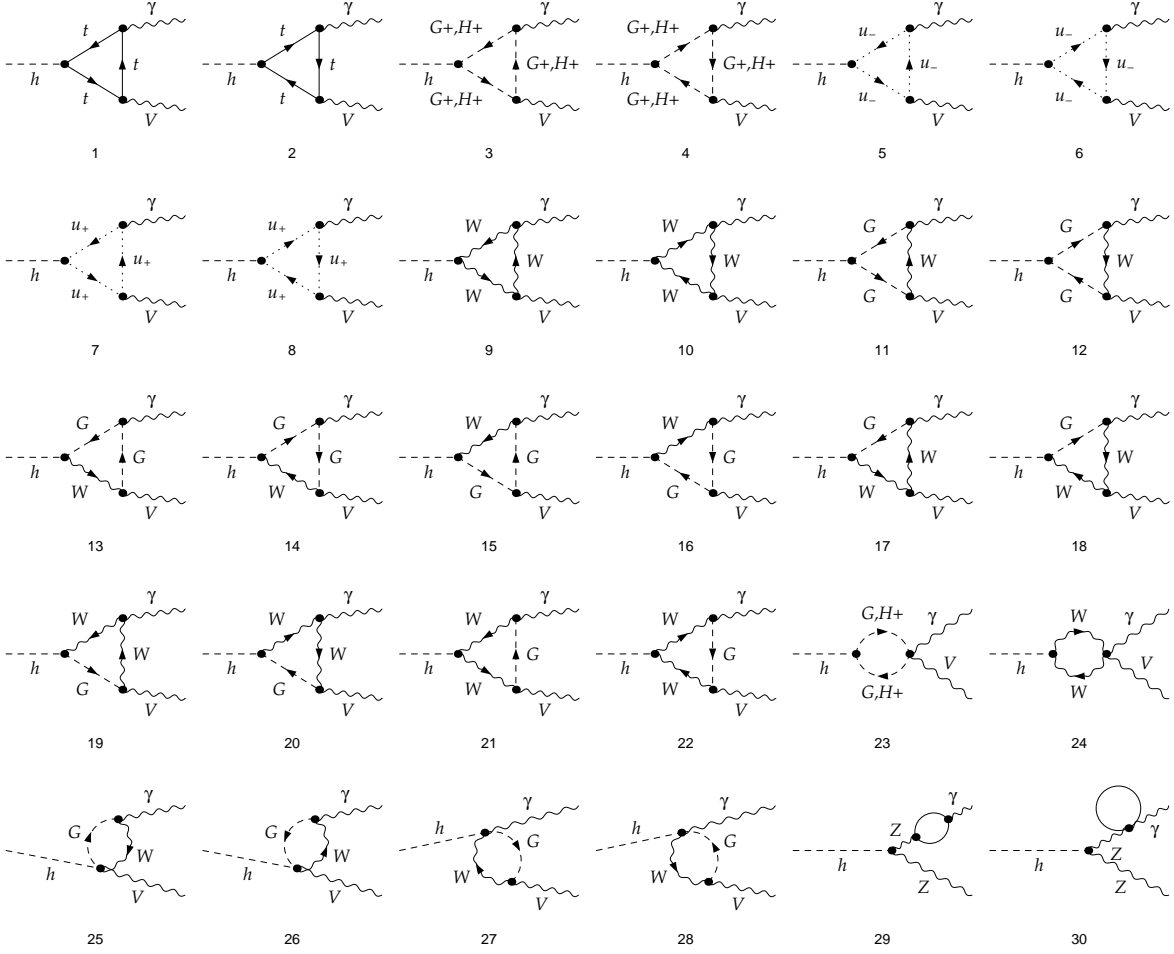}
\end{center}
\caption{Feynman diagrams for $h\to \gamma V$, $V=\gamma$ or $Z$ in the Feynman
gauge. Here $t$, $W^\pm$, $G^\pm$, $u^\pm$, and $H^\pm$ denote 
respectively the top quark, the charged gauge bosons, the Goldstone bosons, 
the Fadeev-Popov ghost and the charged Higgs. }
\label{fih-hgg}
\end{figure}

Recently, many theoretical works have been devoted to the correlation of the 
signal strengths 
between $h \to \gamma \gamma$ and $h \to Z \gamma$ in various models,
in particular the triplet Higgs models \cite{triplet1,triplet2,triplet3} and
two Higgs doublet models \cite{2hdm1,2hdm2}. 
In this section, we will study this correlation in IHDM.
In both processes, one has the same set of charged particles 
circulating the corresponding  loop amplitudes.  
Feynman diagrams contributing to both $h\to \gamma\gamma$ and  
$h\to Z\gamma$ are depicted in Fig.~\ref{fih-hgg} where all particles
inside the loops are shown, with $t$, $W^\pm$, $G^\pm$, $u^\pm$, and $H^\pm$ are 
the top quark, the charged gauge bosons, the Goldstone bosons, 
the Fadeev-Popov ghosts and the charged Higgs bosons respectively. Note that in the case of $h\to Z \gamma$ 
the $\gamma-Z$ mixing as depicted generically in  the last two diagrams (labelled 29 and 30) of 
Fig.~\ref{fih-hgg} has to be taken into account. Only after inclusion of
such mixing that the amplitude for $h\to Z \gamma$ becomes ultra-violet finite. 
%Note that in the Feynman diagrams Fig.~\ref{fih-hgg} (29-30), all  SM charged
%particles as well as the charged Higgs have to be included.
%
Although the final state kinematics as well as 
the $\gamma$ and $Z$ couplings
to the charged particles are different, these 2 channels 
should be correlated to certain extent. 
Compared with SM, the loop amplitudes for the two processes 
receive an additional contribution from the charged Higgs boson 
resided in the inert doublet.
The partial decay width of $h\to\gamma\gamma$ can be found 
in \cite{inert}, while the one for  $h\to Z\gamma$ it is given by
\begin{eqnarray}
&&\Gamma\, (h\to Z\gamma)  =  \frac{G^2_{F}m_W^2\, s_W^2\, \alpha\,m_{h}^{3}}
{64\,\pi^{4}} \left(1-\frac{m_Z^2}{m_{h}^2} \right)^3\bigg|
-2 \frac{(3-8 s_W^2)}{3 s_W c_W} 
\left( I_1(\tau_t, \lambda_t) - I_2(\tau_t,\lambda_t) \right)
\nonumber \\
& &  -\frac{c_W}{s_W} \left[ 4 \left( 3-\frac{s_W^2}{c_W^2} \right) I_2(\tau_W,\lambda_W)
 + \left( \left( 1+\frac{2}{c_W} \right) \frac{s_W^2}{c_W^2} - \left( 5+\frac{2}{c_W}\right) \right) I_1(\tau_W,\lambda_W) \right]
\nonumber\\
&& +\frac{(1-2 s_W^2)}{s_W c_W}\frac{(m_{H^\pm}^2-\mu_2^2)}{m_{H^\pm}^2}
I_1(\tau_{H^\pm},\lambda_{H^\pm})  \bigg|^2
\label{htogaz}
\end{eqnarray}
where $\tau_i= 4m_i^2/m_h^2$ and $\lambda_i= 4m_i^2/m_Z^2$ $(i=t,W^\pm,H^\pm)$. 
We only show the most dominant top quark contribution in the fermion loops.
The loop functions $I_1$ and $I_2$ can be found in the 
literature~\cite{HiggsHunterGuide}. 
Recall that in the SM, the decay widths of the two processes 
are dominated by the $W$ loop which interferes {\it destructively} with the subdominant top quark 
contribution. 
The extra charged Higgs contribution is shown in the last term of 
Eq.~(\ref{htogaz})
which is proportional to the SM Higgs coupling to a pair of $H^\pm$,
\begin{equation}
g_{hH^{\pm}H^{\mp}}=-2\,\frac{m_W s_W}{e}\lambda_3 = 
\frac{e(m_{H^\pm}^2-\mu_2^2)}{2 m_W s_W} \; .
\label{inerthp}
\end{equation}
It is clear from the above Eq.~(\ref{inerthp}) that the coupling
$g_{hH^{\pm}H^{\mp}}$ is completely fixed by the parameter
$\lambda_3$.  Just like the case of $h\to\gamma\gamma$ \cite{inert}, 
for negative and positive $\lambda_3$, charged Higgs contribution
can enhance and suppress the $h\to Z\gamma$  rate respectively. 
A preliminary result for such correlation between $h\to\gamma\gamma$
and $h\to Z\gamma$ in the IHDM was first presented\footnote{In this paper,
due to a bug in our old code some plots for $h\to Z \gamma$ and its correlation 
with $h\to \gamma\gamma$ are slightly modified as compared to one presented in
\cite{aa-hapnp}. The results now are in agreement with
\cite{Swiezewska:2012eh}.} in \cite{aa-hapnp}.  
Recently this correlation has been discussed in 
\cite{Swiezewska:2012eh,marianew} and similar results were found.

%
%=======================================
The largest contribution to the production cross section of the Higgs 
is through gluon fusion. For the Higgs decays to $\gamma\gamma$ or 
$\gamma Z$ channels, one defines the signal strength as 
the ratio of production cross section times branching ratio 
normalized to the SM one as
\begin{eqnarray}
R_{\gamma V}& =& \frac{\sigma(gg\to V\gamma )}{\sigma(gg\to V \gamma)^{SM} } \approx
\frac{\sigma(gg\to h)\times Br(h \to V\gamma) }{\sigma(gg\to
  h)^{SM}\times Br(h \to V\gamma)^{SM} }  \quad , \quad V=(\gamma , Z)
\label{ratioVG}
\end{eqnarray}
where the narrow width approximation has been used.
Since the Higgs $h$ has the same couplings to 
fermions in IHDM as in SM, the corresponding production cross sections from gluon fusion 
are the same and Eq.~(\ref{ratioVG}) reduces to just the ratio of branching ratios. 
Moreover, if the invisible decay $h\to \chi \chi$ is not open, the total
widths of the Higgs in both models will be approximately the same
and Eq.~(\ref{ratioVG}) will further reduce to the ratio 
of the partial widths of $h\to V\gamma$  in both models.
In our numerical work, we perform a systematic scan over all the allowed parameter space $\cal P$ 
with $m_h$ set at 125 GeV, 
taking into account all the theoretical and experimental constraints described 
in previous section.
We note that once the invisible decay $h \to \chi \chi$ is open, 
its branching ratio will dominate over all other SM channels unless one 
tunes the coupling $g_{h\chi\chi} = - 2(m_\chi^2-\mu_2^2)/v$ to be very small 
by taking $m_\chi^2\approx \mu_2^2$.
The opening of the invisible mode of $h$ would enhance 
the total width of the SM Higgs boson and therefore suppress both 
$h\to \gamma\gamma$ and $h\to Z\gamma$ branching ratios.
We will consider the case where $m_h < 2 m_\chi$
such that the invisible mode of $h \to \chi \chi$ is close and
Eq.~(\ref{ratioVG}) reduces to just ratio 
of partial widths as mentioned above.
%
%%%%%%%%%%%%%%%%
\begin{figure}[t]
\begin{centering}
\includegraphics[height = 2.8in, width=2.5in]{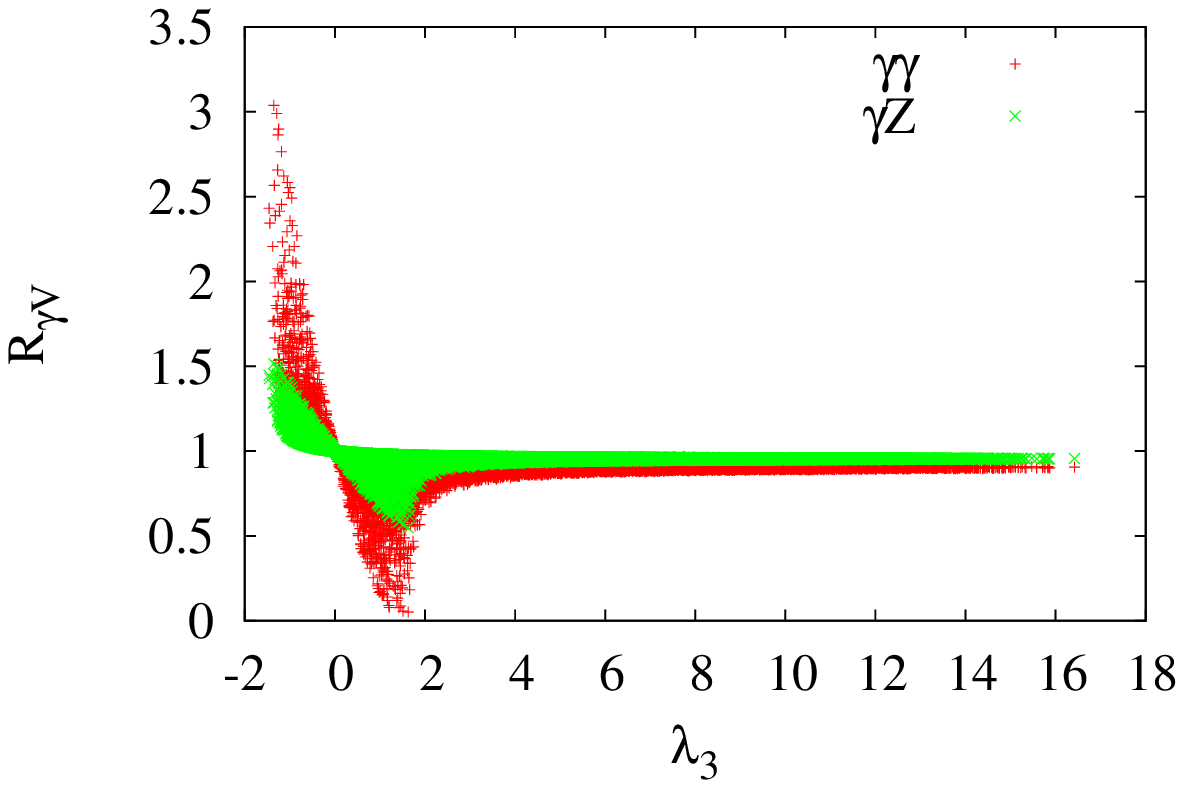}
{\hspace{-1.5cm}\includegraphics[height = 2.8in, width=2.5in]{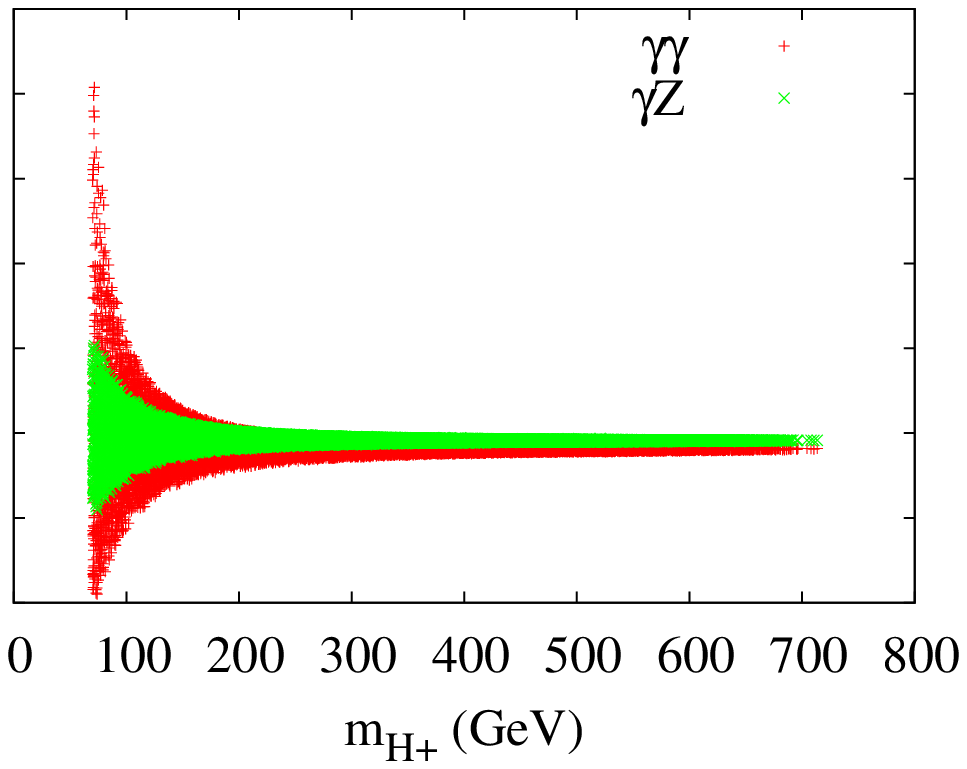}}
{\hspace{-1.5cm}\includegraphics[height = 2.8in,width=2.5in]{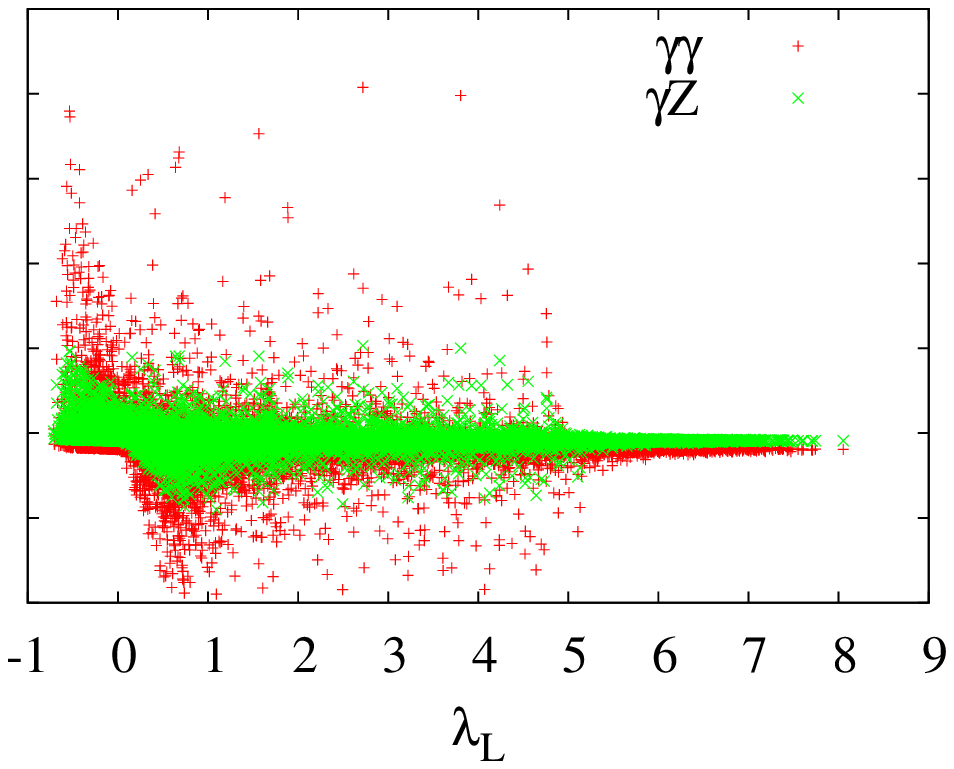}}
\end{centering}
\caption{Signal strength $R_{\gamma V}$ as function of
$\lambda_3$ ({\it left}), $m_{H^\pm}$ ({\it middle}) and $\lambda_L$ 
({\it right}). We scan over $m_{H^\pm}\in [70, 500]$ GeV, $-10^6\leq \mu_2^2\leq
  10^6$ GeV$^2$ and $0\leq \lambda_2\leq 4\pi/3$.
} 
\label{fig1}
\end{figure}
%%%%%%%%%%%%%%%%%%%%%%%%%%
\begin{figure}[t]
\begin{center}
\includegraphics[height = 3.0in,width=3.25in]{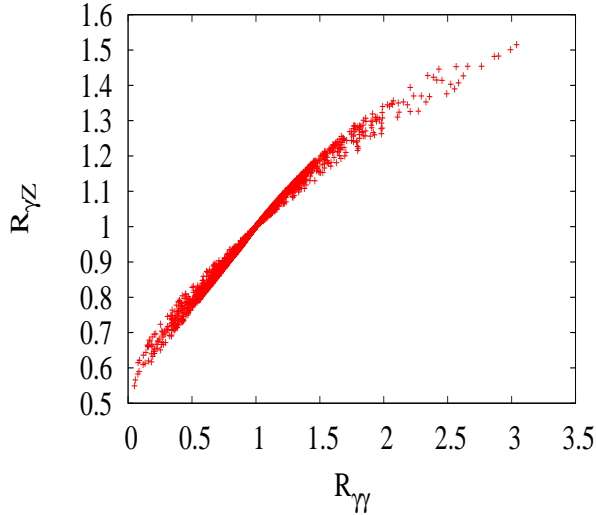}
\end{center}
\vspace{-1.5cm}
\caption{Correlation between $R_{\gamma\gamma}$  and $R_{\gamma Z}$ in IHDM.  
Parameter scan same as Fig.~\ref{fig1}.
} 
\label{fig2}
\end{figure}
%%%%%%%%%%%%%%%%%%%%%%%%%%%%%%%%%%
%

Results of our scans are depicted in Figs.~\ref{fig1} and \ref{fig2}.
In Fig.~\ref{fig1}, we plot $R_{\gamma V}$ as a 
function of $\lambda_3$ (left), $m_{H^\pm}$ (middle)
and $\lambda_L$ (right). 
From the first and second plots, it is clear that to enhance substantially
$R_{\gamma\gamma}$ and $R_{\gamma Z}$ we need a negative $\lambda_3$ and 
a rather light charged Higgs. 
The enhancement in $h\to Z\gamma$  is always smaller than  
in $h\to \gamma\gamma$ because the coupling ratio 
$g_{Z H^\pm H^\mp} / g_{\gamma H^\pm H^\mp} =(1-2 s_W^2)/(2
s_Wc_W)\approx 0.67 $. The suppression factor 
of $R_{\gamma Z}$ versus $R_{\gamma\gamma}$ is therefore {$(0.67)^2$}.
The lighter the charged Higgs is, the more pronounced in the enhancement 
of the $\gamma \gamma$  and $Z\gamma$ rates. 
For instance, if we need $R_{\gamma\gamma}\geq 1.1$ or $R_{\gamma Z}\geq 1.1$
for $\lambda_3<0$ ({\it i.e.} $\mu_2^2> m_{H^\pm}^2$),
the charged Higgs mass $m_{H^\pm}$ has to be lighter than 200 or 115 GeV respectively.
In the rightmost of Fig.~\ref{fig1}, we plot $R_{\gamma V}$ as a
function of $\lambda_L$ in the range of $[-2,2]$. We note that
$\lambda_L$  is an important parameter which enters in the calculation of the relic density of DM
and hence it is constrained by the WMAP data  to be in the range of 
$|\lambda_L|<0.2$  \cite{Dolle:2009fn}.  
More sophisticated limits of this parameter $\lambda_L$ depending on the mass $m_\chi$ have been
deduced recently from a global analysis of IHDM \cite{Arhrib:2013ela}.
We note that the stringent limits obtained in \cite{Arhrib:2013ela} all lie within the range of $[-2,2]$,
thus some enhancements in $R_{\gamma\gamma}$ and $R_{\gamma Z}$ are still possible according to this plot.
The correlation between  $R_{\gamma\gamma}$ and  $R_{\gamma Z}$ 
is  roughly a linear one as shown in Fig.~\ref{fig2} 
using the same parameter scan as in Fig.~\ref{fig1}. From the plot, one finds that for $R_{\gamma\gamma}>1$ 
where the $W^\pm$ and $H^\pm$ loops interfere constructively,
we have $R_{\gamma\gamma}\geq R_{\gamma Z}$;  
while in the opposite case of $R_{\gamma\gamma}<1$ where the
$W^\pm$ and $H^\pm$ loops interfere destructively, we can have 
$R_{\gamma\gamma} \leq R_{\gamma Z}$. The main reason for this feature is that 
the destructive interference between $W^\pm$ and $H^\pm$ 
is more effective in $R_{\gamma\gamma}$ than in $R_{\gamma Z}$ since the latter process 
has a much larger $W^\pm$ contribution.
%
%%%%%%%%%%%%%%%%%%%%%%%%%%%%%%%
%
\begin{figure}[t]
\begin{center}
\includegraphics[scale=1.03]{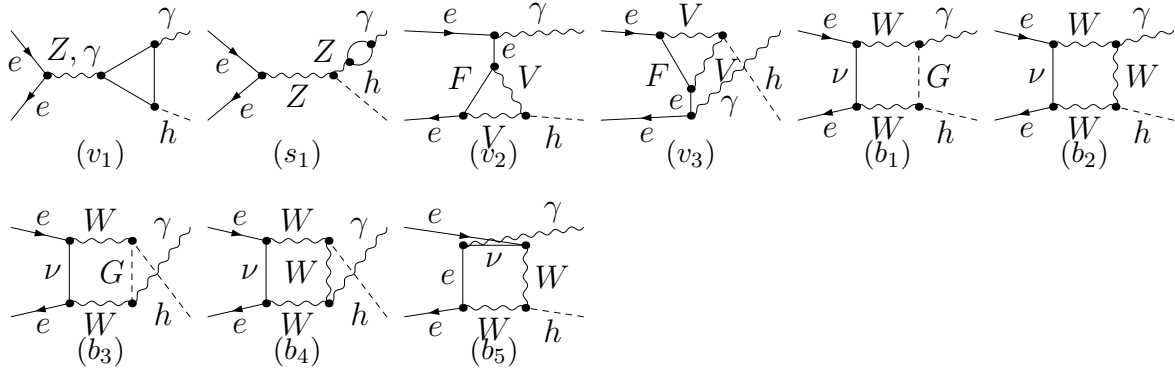}
\end{center}
\caption{ Generic Feynman diagrams contributing to $e^-e^+\to h\gamma$.
For the particles inside the loops in diagrams $v_1$, 
we have all possible charged particles given in Fig.~\ref{fih-hgg}. Mixing
$\gamma-Z$ diagram $s_1$ receives contributions from all SM 
particles as well as charged Higgs.
For diagrams $v_{2,3}$, $V$ denotes $Z$ or $W$, while $F$ denotes $e$ or $\nu$. 
Box diagrams $b_{1,...,5}$ are necessary for gauge invariance.}
\label{fig3}
\end{figure}

%%%%%%%%%%%%%%%%%%%%%%%%%%%%%%
%==============================================================
\section{Associated Production $e^+e^- \to \gamma h$ and $e^- \gamma \to e^- h$ in IHDM}
\begin{figure}[b]
\begin{center}
\includegraphics[scale=1.03]{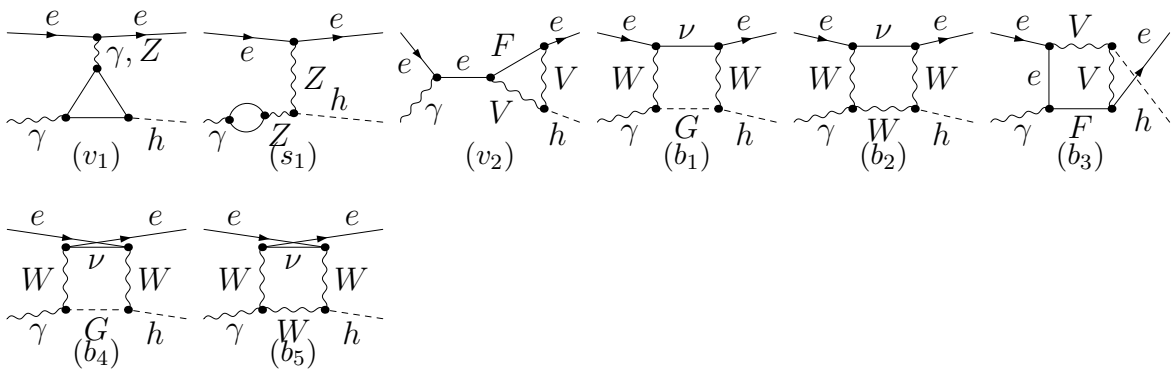}
\end{center}
\caption{Generic Feynman diagrams contributing to $e^-\gamma\to e^-h$.
For the particles inside the loops in diagrams $v_1$, 
we have all possible charged particles like in Fig.\ref{fih-hgg}. Mixing
$\gamma-Z$ diagram $s_1$ receives contributions from all SM 
particles as well as charged Higgs.
For diagram $v_{2}$, $V$ denotes $Z$ or $W$, while $F$ denotes $e$ or $\nu$. 
Box diagrams $b_{1,...,5}$ are necessary for gauge invariance.
}
\label{fig4}
\end{figure}

%==============================================================
At tree level, the associated production process of $e^+e^- \to \gamma h$ 
is mediated by t-channel electron exchange diagram which is suppressed by the
electron mass. For the process $e^- \gamma \to e^- h$,
 the tree level contribution is mediated by s-channel diagrams which is also  
 suppressed by the electron mass.
At one-loop level, they are mediated by triangle, 
self-energy as well as box diagrams
and hence they are sensitive to all virtual particles 
(physical gauge bosons, fermions and charged Higgs particles as well as 
unphysical Goldstone $G^\pm$ and ghost particles $u^\pm$) inside the loop.
We display in Figs.~(\ref{fig3})  and (\ref{fig4}) some generic 
Feynman diagrams that contribute to $e^+e^- \to \gamma h$ and 
$e^- \gamma \to e^- h$ respectively, indicating that some individual amplitudes
are sensitive to the off-shell $h\gamma V^*$ vertices. 
Both in Figs.~(\ref{fig3})  and (\ref{fig4}), 
diagrams $v_1$ are generic one
and the particles content is depicted in Fig.~\ref{fih-hgg}.
The process $e^+e^-\to \gamma h$ had been studied in SM 
long time ago \cite{Barroso:1985et,Abbasabadi:1995rc}. 
Effects from new physics to this process 
had been analyzed in Ref.~\cite{Djouadi:1996ws} for supersymmetry 
and Ref.~\cite{Akeroyd:1999gu} for an extended Higgs sector.
%
%%%%%%%%%%%%%%%%%%%%%%%%%%%%%%
%%%%%%%%%%%%%%%%%%%%%%%%%%%%%%
Our calculation is done in Feynman gauge using dimensional regularization 
with the help of {\sl FeynArts} and {\sl FormCalc} packages \cite{FA2}. 
Numerical evaluation of the scalar integrals is done with 
{\sl LoopTools} \cite{FF}. 
Throughout the calculation we will neglect the electron mass.
Since the tree level amplitudes which are 
suppressed by the electron mass are neglected,
Feynman diagrams like Fig.~{\ref{fig3}}-$v_2$, $v_3$ and Fig.~{\ref{fig4}}-$v_2$ 
are ultraviolet finite because the corresponding counter-terms for $e^+e^-h$ 
are proportional to electron mass.
Note also that in the on-shell renormalization scheme such as Ref.~\cite{A-H},
there are no counter-terms for $\gamma\gamma h$ and $Z\gamma h$.
We have checked both analytically and numerically 
that the total amplitudes for the two processes are ultraviolet finite. 
The $\gamma-Z$ self-energy mixing is necessary in order to achieve the finite results. 
While the fermionic contributions to $v_1$ (triangle) 
and $s_1$ (self-energy) diagrams in 
Figs.~(\ref{fig3}) and (\ref{fig4}) are gauge invariance by themselves, 
for gauge boson diagrams 
we need to sum these with all other (triangle and box) 
diagrams in order to maintain gauge invariance in the final 
results \cite{Djouadi:1996ws}. 
In all Feynman diagrams computed here, there 
is no virtual photon in the loops, therefore the results are infrared finite.
Real or virtual emission of the photon is suppressed by the electron mass. 
For illustrative purpose in the following, it is convenient to introduce 
the two ratios
\begin{eqnarray}
R_{\gamma h}  \equiv  \frac{\sigma(e^+e^-\to \gamma h)}{
\sigma_{\rm SM}(e^+e^- \to \gamma h)} \;\;\; , \;\; 
R_{e^- h}  \equiv  \frac{\sigma(e^- \gamma \to e^- h)}{\sigma_{\rm SM}(e^- \gamma\to e^-h)} \; ,
\end{eqnarray}
which are the total cross sections in the IHDM normalized to the SM one.
%==========================
\begin{figure}[t]
\bcen
\epsfig{file=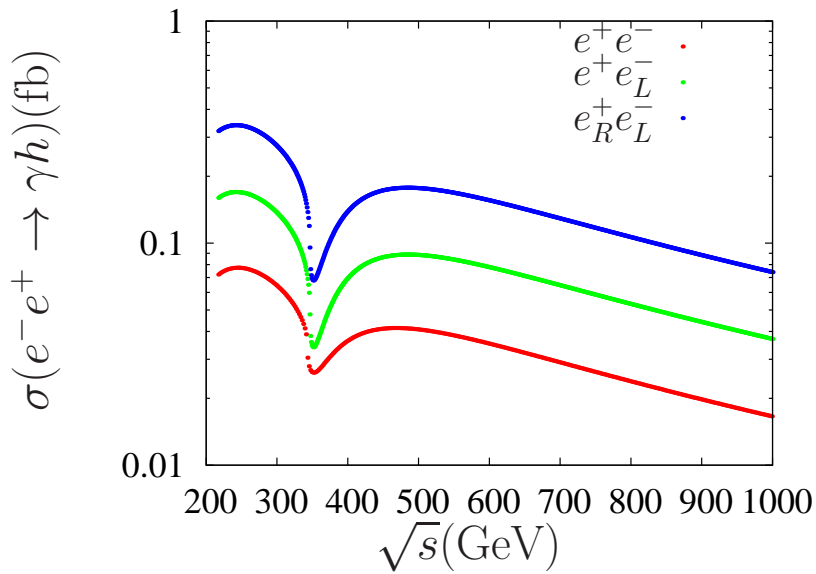,width=0.7\textwidth}
\caption{Total cross section for $e^+e^-\to \gamma h$ (fb) in the SM as a function of
center-of-mass energy with $m_h=125$ GeV. From bottom to top: unpolarized,
left polarized electron, left polarized electron and right polarized positron. 
}
\label{fig5}
\ecen
\end{figure}
%%%%%%%%%%%%%%%%%%%%%%%%%%%%%
%%%%%%%%%%%%%%%%%%%%%%%%%%%%%
\begin{figure}[b]
\bcen
\epsfig{file=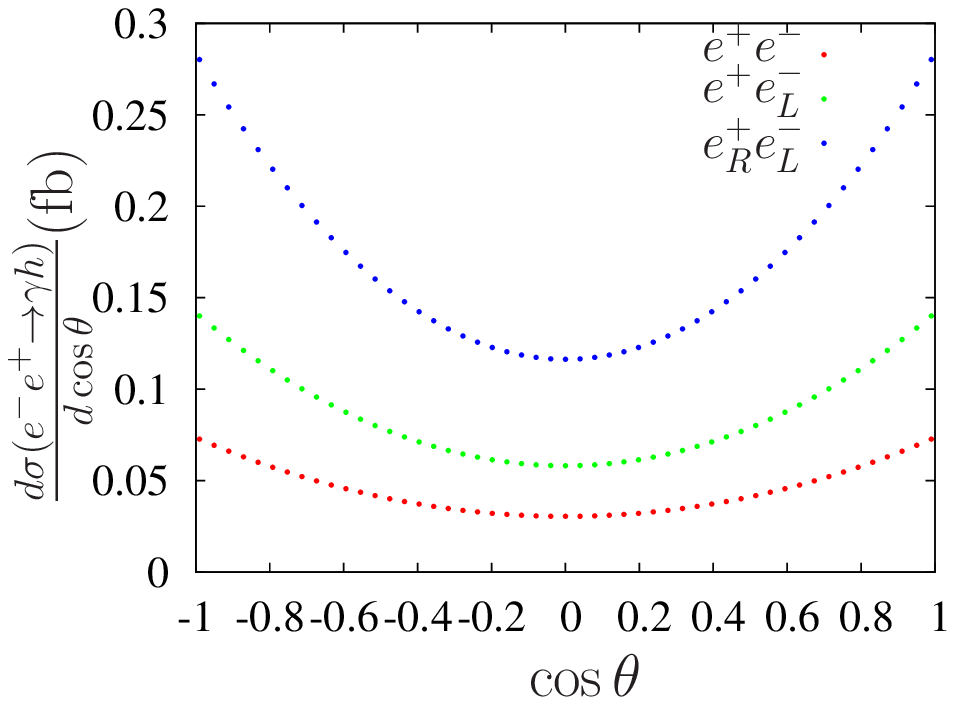,width=0.55\textwidth}
\hspace*{-2cm}
\epsfig{file=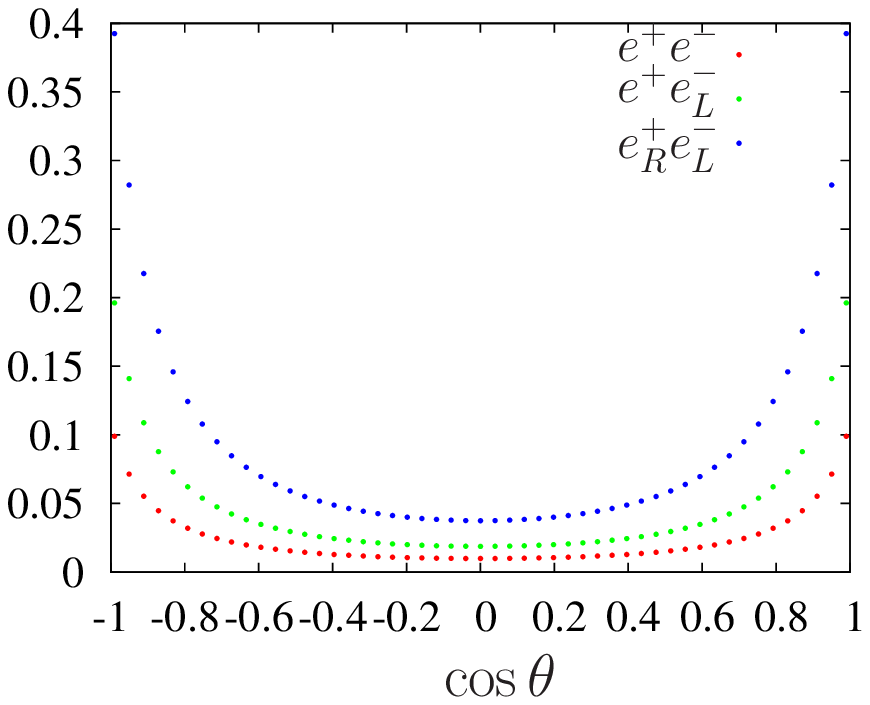,width=0.55\textwidth}
\caption{Differential cross section for $e^+e^-\to \gamma h$ (fb) in the SM
with $m_h=125$ GeV and $\sqrt{s}$ = 250 ({\it left}) and 
500 ({\it right}) GeV.
}
\label{fig6}
\ecen
\end{figure}
%%%%%%%%%%%%%%%%%%
\subsection{$e^+ e^- \to \gamma h$}
In Fig.~\ref{fig5}, we plot the associated production cross section of the 125 GeV SM 
Higgs with a photon at the linear collider as a function of center-of-mass energy $\sqrt s$
from 200 GeV to 1 TeV.
The lower, middle and upper curves correspond to the unpolarized $e^+e^-$,
polarized $e^+ e^-_L$ and $e^+_R e^-_L$ beams respectively.
In all three cases, the cross sections are enhanced near the 
region of $\sqrt{s}\approx 250$ GeV. As the center-of-mass energy increases further, the destructive interference between 
the top quark and $W^\pm$ contributions get more severe 
and become maximal near the $t\bar{t}$ threshold, responsible for the dips clearly seen in the figure.
After crossing the $t\bar{t}$ threshold, the cross sections scale like $1/s$ and thus drop steeply. 
Note that with the polarization of the initial state of positron or both electron and positron the total cross section 
can be increased by roughly a factor 2 or 4 respectively compared with the unpolarized case.
In Fig.~\ref{fig6} we exhibit the corresponding angular distribution $d\sigma/d\cos\theta$ 
at $\sqrt{s}$ =250 (left) and 500 (right) GeV respectively. One observes that at both energies 
the distributions are rather symmetric for either choices of  polarized or unpolarized beams.
%==============================================================
%%%%%%%%%%%%%%%%%%
\begin{figure}[!]
\bcen
\epsfig{file=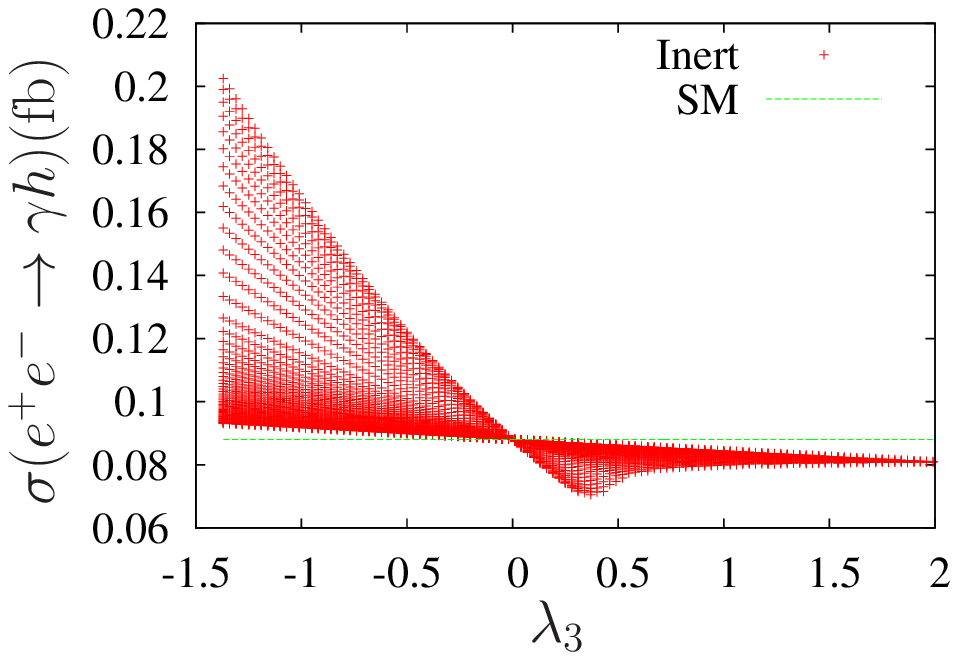,width=0.51\textwidth}
\hspace*{-.6cm}
\epsfig{file=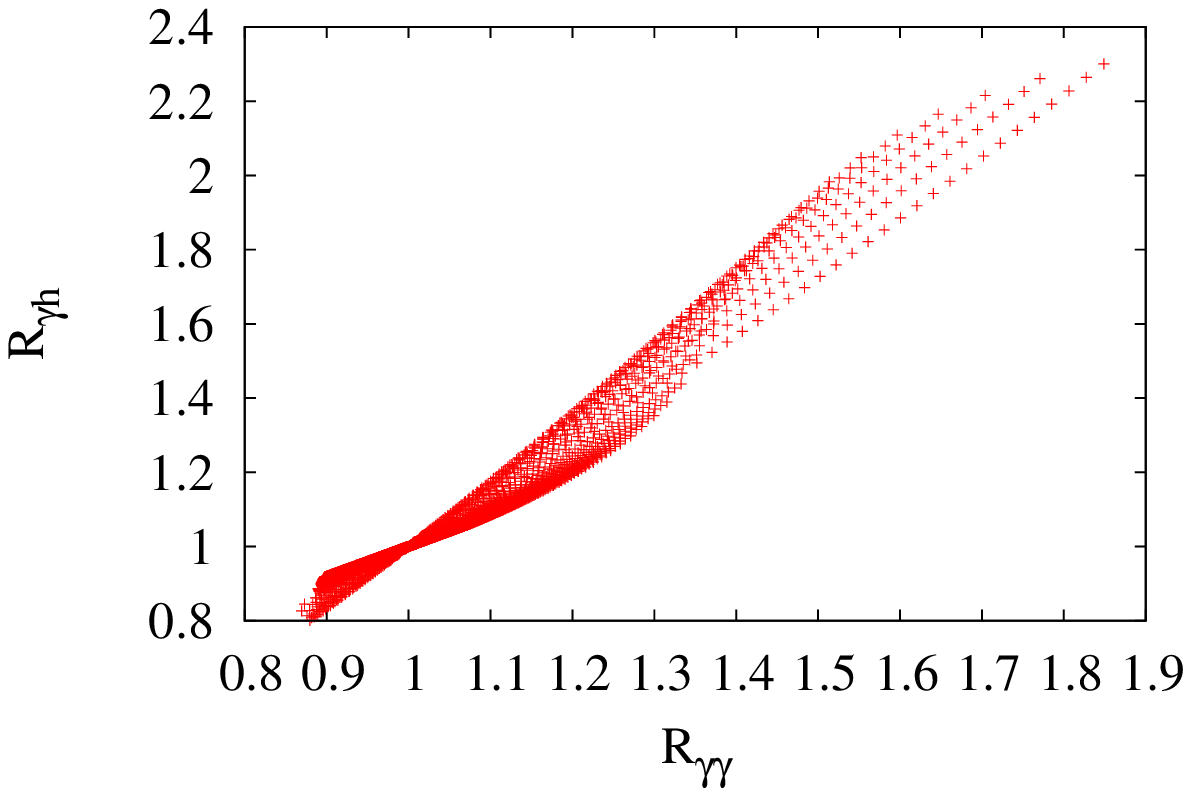,width=0.51\textwidth}
\epsfig{file=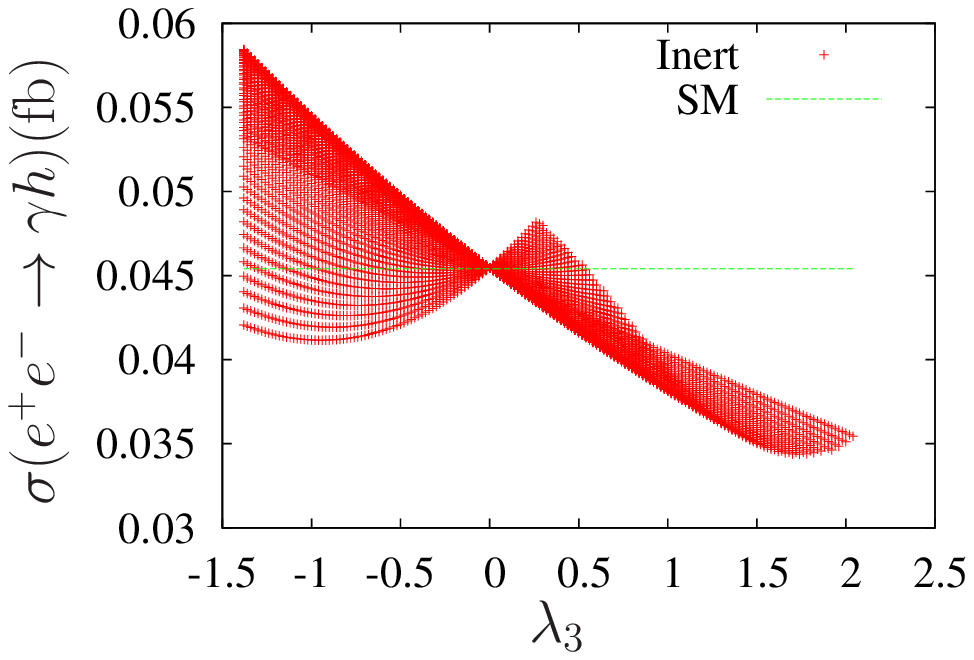,width=0.51\textwidth}
\hspace*{-.6cm}
\epsfig{file=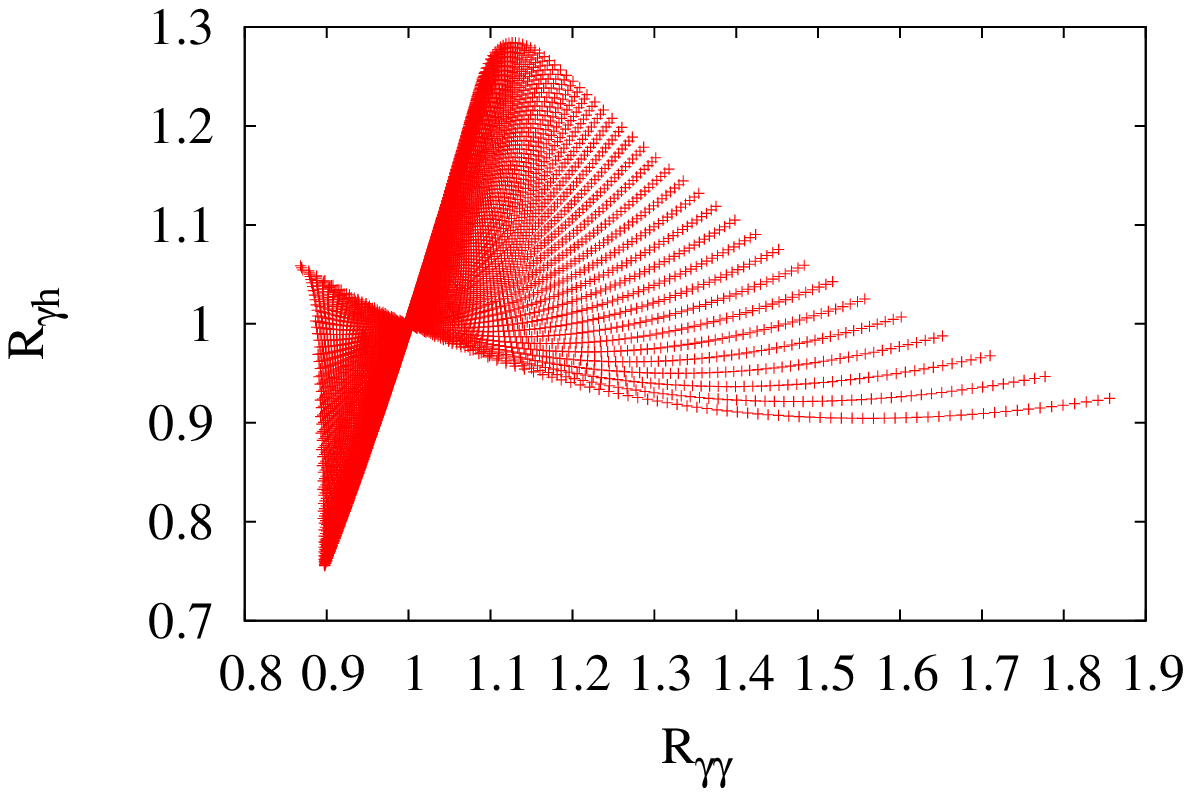,width=0.51\textwidth}
\caption{Total cross section for $e^+e^- \to \gamma h$ in IHDM as a function of $\lambda_3$ ({\it left}) 
and correlation between $R_{\gamma\gamma}$ and $R_{\gamma h}$ ({\it right}).
Input parameters are $m_h=125$ GeV, $\lambda_2=3.75$, $\mu^2_2 \in [0, 5]$ TeV$^2$, and 
$m_A=m_S=m_{H^\pm}+10$ GeV with $m_{H^\pm}\in [90,350]$ GeV.
Upper and lower plots correspond to $\sqrt{s}=250$  
and 500 GeV respectively.
} 
\label{fig7}
\ecen
\end{figure}
%%#######################################################%%
In Fig.~\ref{fig7} we show the total cross section \eegah 
as a function of $\lambda_3$ (left) and 
the correlation between $R_{\gamma\gamma}$ and $R_{\gamma h}$ (right)
for $\lambda_2=3.75$ and $m_A=m_S=m_{H^\pm}+10$ GeV with $m_{H^\pm} \in [90,350]$ GeV. 
Upper and lower plots are for $\sqrt{s}$ = 250 and 500 GeV respectively.

For $\sqrt{s}$ = 250 GeV (two upper plots in Fig.~\ref{fig7}), it is clear that when $\lambda_3$ is negative, 
interference of $H^\pm$ and $W^\pm$ loops in the off-shell $h\gamma V^* (V = \gamma, Z)$ 
amplitudes is also constructive and it can give rise to some enhancement in the total cross section 
of \eegah with respect to SM.  The increase can be as large as a factor of 2. 
This large enhancement requires of course a rather light charged Higgs in the range
$[90,200]$ GeV circulating inside the loop.
Given the fact that the enhancement of the cross section happens for negative $\lambda_3$ which is the same 
condition for having an enhancement in $R_{\gamma\gamma}$ and $R_{\gamma Z}$, 
the correlation between $R_{\gamma\gamma}$ and $R_{\gamma h}$ is shown 
in the upper right plot of Fig.~\ref{fig7}.
Clearly, when  $R_{\gamma\gamma}>1$ we also have $R_{\gamma h}>1$.
Similar correlation can be found between $R_{\gamma Z}$ and $R_{\gamma h}$ 
but will not be shown here.

For $\sqrt{s}=500$ GeV (two lower plots in Fig.~\ref{fig7}), the top quark contribution gets amplified after crossing 
the $t\bar{t}$ threshold which leads to more destructive interference 
with the $W^\pm$ loops as can be seen in the lower left plot in Fig.~\ref{fig7}  
for both positive and negative $\lambda_3$.
At this higher energy, the destructive top quark loop can overwhelm the constructive interference between 
the $H^\pm$ and $W^\pm$ loops with negative  $\lambda_3$ such that the total cross section 
is below its SM value. The opposite case is 
also possible, the top quark loop can be constructive interference with the $H^\pm$ loop
for positive  $\lambda_3$ and overwhelm the $W$ loop leading to a total cross section 
larger than its SM value. In the lower right plot of In Fig.~\ref{fig7}, the correlation between 
$R_{\gamma\gamma}$ and $R_{\gamma h}$ is also shown for $\sqrt{s}$ = 500 GeV.
%%%%%%%%%%%%%%%%%%%%%%%%%%
\begin{figure}[htb]
\bcen
\epsfig{file=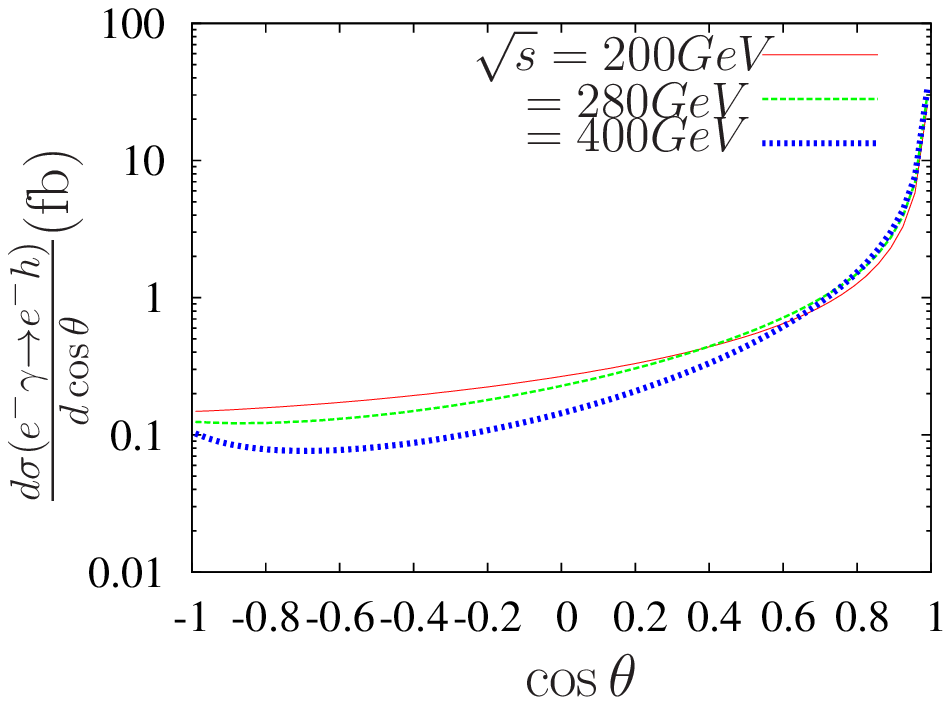,width=0.52\textwidth}
\hspace*{-.9cm}
\epsfig{file=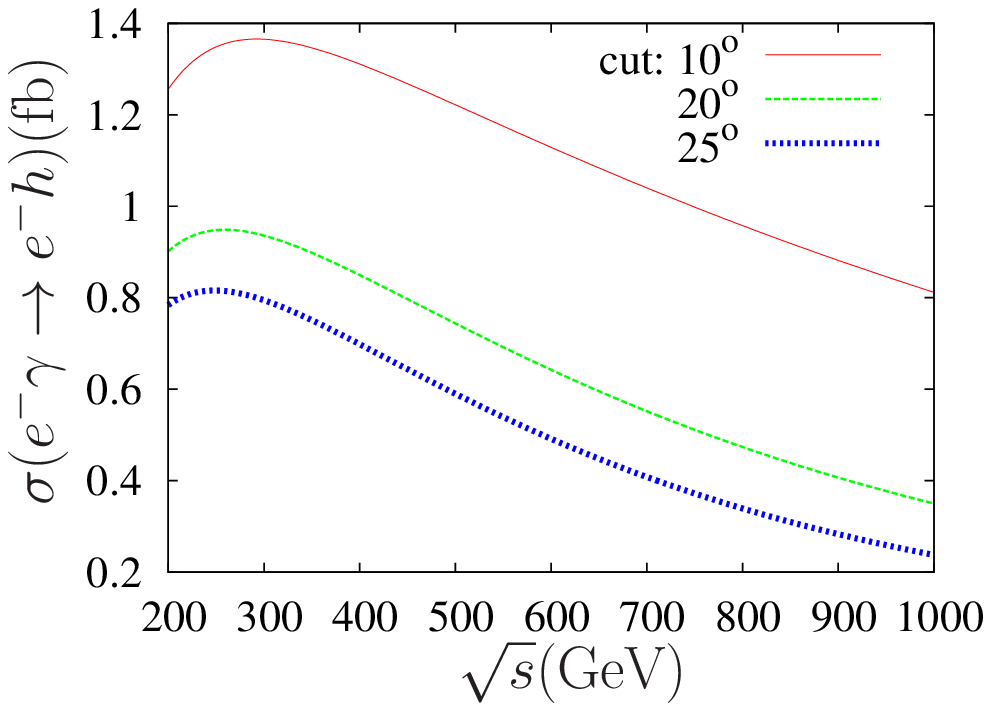,width=0.52\textwidth}
\caption{({\it Left}) SM differential cross section for $e^-\gamma\to e^- h$ (fb) 
with $m_h=125$ GeV for $\sqrt{s}$ = 200 (red), 280 (green) and 400 (blue) GeV.
({\it Right}) SM total cross section as a function of center-of-mass energy for 3 different 
angle cuts of $10^\circ$ (red), $20^\circ$ (green) and $25^\circ$ (blue).
} 
\label{fig8}
\ecen
\end{figure}
%%%%%%%%%%%%%%%%%%%%%%%%%%%%%%%%
\subsection{$e^- \gamma \to e^- h$}
Next we turn to discuss the case of $e^-\gamma \to e^- h$.
If the $e^-\gamma$ option for ILC is made available, the center-of-mass
energy may be slightly reduced as compared with the previous $e^+e^-$ case
shown in Fig.~\ref{fig6}. 
In the left plot in Fig.~\ref{fig8}, we show the differential cross section 
$d \sigma (e^-\gamma \to e^- h)/ d \cos \theta$
for three center-of-mass energies at $\sqrt{s}$ = 200, 280 and 400 GeV.
Obviously, the differential cross section for this $e^- \gamma$ case get significantly enhanced 
near the forward direction $\cos\theta\approx 1$
due to the $t$ channel singularity between the incoming and outgoing 
electrons. To avoid this kinematical singularity we will impose a cut on the scattering angle
when computing the total cross section.
In the right plot in Fig.~\ref{fig8}, we illustrate the total cross section
$\sigma (e^-\gamma \to e^- h)$ as a function of center-of-mass energy 
for three different cuts of 10$^\circ$, 20$^\circ$ and 30$^\circ$ on the scattering angle. 
The sensitivity to the angular cut is quite evident.
At $\sqrt{s}\approx 250$ GeV, the total cross section can reach a maximum value of 1.38,
0.95 and 0.8 fb for 10$^\circ$, 20$^\circ$ and 30$^\circ$ cuts respectively. 
We will use a 20$^\circ$ cut on the scattering angle in our next figure.
%%%%%%%%%%%%%%%%%%%%%%%%%%%%%%%%%%%%%
\begin{figure}[htb]
\bcen
\hspace*{-0.8cm}
\epsfig{file=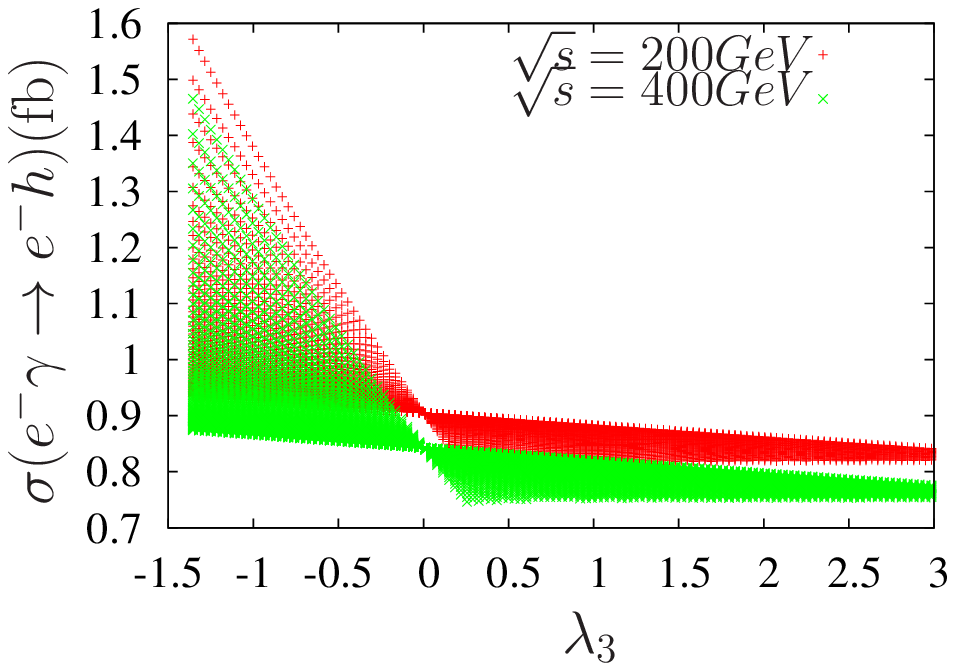,width=0.52\textwidth}
\hspace*{-0.8cm}
\epsfig{file=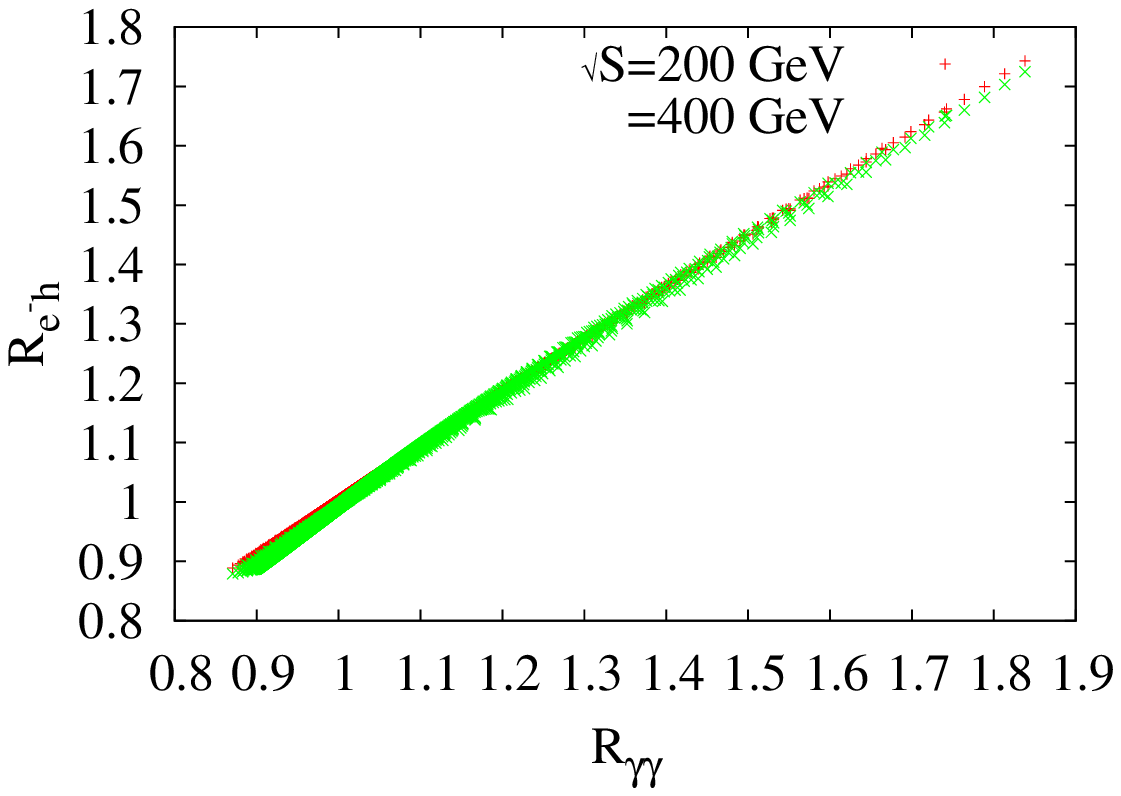,width=0.52\textwidth}
\caption{Total cross section for $e^-\gamma\to e^- h$ (fb) in IHDM as a function of $\lambda_3$ ({\it left})
and correlation between $R_{\gamma\gamma}$ and $R_{e^- h}$  ({\it right}) for 
$\sqrt{s}$ = 200 (red)  and 400 (green) GeV.
Parameter scan same as Fig.~\ref{fig7}.} 
\label{fig9}
\ecen
\end{figure}
%%%%%%%%%%%%%%%%%%%%%%%%%%%%%%%%%%%%%%%%
In the left plot of Fig.~\ref{fig9} we show the total cross section for  
$e^-\gamma \to e^- h$ as a function of $\lambda_3$ for $\sqrt{s}=200$ and
$400$ GeV. It is clear that the charged Higgs loop  
interferes constructively (destructive) with the SM loops for negative (positive) $\lambda_3$
respectively. The lighter the charged Higgs mass is, the larger 
the enhancement in the total cross section $\sigma(e^-\gamma \to e^- h)$. 
In the right plot of Fig.~\ref{fig9}, the correlation between $R_{\gamma\gamma}$ 
and $R_{e^-h}$ is shown for $\sqrt{s}$ = 200 and 400 GeV. At both energies, 
we have a positive correlation and $R_{\gamma\gamma} \approx R_{e^-h}$.
Similar conclusion can be drawn for the correlation between 
$R_{\gamma Z}$ and $R_{e^-h}$ which we will not show it here.

\section{Conclusions}

Despite the discovery of the SM Higgs, the search for physics beyond the 
SM is just getting started. 
Since all the current LHC and Tevatron data point toward this new boson is indeed the SM Higgs 
with its couplings consistent with the SM expectations, it leaves little room for an extra Higgs
doublet to play any role in the spontaneous symmetry breaking.
In this regard, IHDM is quite an interesting model beyond the SM since its
extra Higgs doublet is inert.

In Section III of this paper we analyzed the 
correlation between the LHC signal strengths $R_{\gamma\gamma}(h)$ and $R_{\gamma Z}(h)$ 
in the IHDM with the possible deviation from their SM predictions of unity.
We have considered the scenario where the invisible modes are not opened.
In this case, we have shown that $R_{\gamma\gamma}(h)$ and $R_{\gamma Z}(h)$ are positively
correlated with a roughly linear relation between the twos. 
Depending on whether 
$R_{\gamma\gamma}(h)$ is greater or lesser than 1, we found that 
$R_{\gamma Z}(h)$ is lesser or greater than $R_{\gamma\gamma}(h)$ respectively,
due to the intricate destructive interference between the contribution of the additional charged Higgs  
with the SM $W^\pm$ inside the loop of the two processes.
While the decay mode $h\to \gamma \gamma$ has played an important role in the discovery of the SM Higgs at the LHC,
the mode $h \to Z \gamma$ has yet to be verified. We expect LHC-14 should be able to detect this latter 
mode positively and provide useful correlation information among these two modes. The correlation 
of these two signal strengths studied in this work for IHDM can then be tested accordingly at the LHC-14.

Due to its clean environment, ILC has great potential 
to measure various properties of the SM Higgs more precisely. 
These include branching ratios, cross sections, CP properties and its mass. 
In Section IV of this paper, we have computed the one-loop processes 
$e^+e^-\to \gamma h$ and $e^-\gamma \to e^- h$ 
in the Feynman gauge using dimensional regularization for the future ILC machine. 
We have shown that the
charged Higgs loops in IHDM can modify the SM predictions for these processes
in a significant way. For both processes,  we have calculated in the IHDM the total 
as well as the differential cross section for the recently discover Higgs at 125 GeV.
We also studied the total cross sections for these two processes 
as a function of the parameter $\lambda_3$ which controls the contribution 
of the charged Higgs boson in the loops. We find that the 
cross sections for both processes are quite sensitive to this parameter 
so that the signal strengths $R_{\gamma h}$ and $R_{e^- h}$ that
we defined for the ILC can be deviated from their SM values of unity.
Furthermore, we have studied the correlation of these two signal strengths
with $R_{\gamma \gamma}$.
We found that for the correlation between $R_{\gamma h}$ and $R_{\gamma \gamma}$ can be mainly positively 
for $\sqrt s$ = 250 GeV and either positive or negative correlated for $\sqrt s$ = 500 GeV depending on the 
IHDM parameter space.
On the other hand, for the correlation between $R_{e^- h}$ and $R_{\gamma \gamma}$, we found a roughly 
linear relation between them for both $\sqrt s$ = 200 and 500 GeV.
All our predictions for the IHDM in this work can be tested at the ILC.

\newpage

\section*{Acknowledgements}
%%#######################################################%%
AA would like to thank the hospitality of Institute of Physics, Academia Sinica in Taiwan 
where progress of this work was made. This work was supported in part by the
Spanish Consejo Superior de Investigaciones Cientificas (CSIC) (RB) and 
by the National Science Council of Taiwan under project number 102-2811-M-001-032 (AA) and 
grant number 101-2112-M-001-005-MY3 (TCY).
%%%%%%%%%%%%%%%%%%%%%%%%%%%%%%%%%%%%%%%%%%%%%%%%%%%%

% References

\end{document}